\documentclass[aps,pra,reprint,showpacs,groupedaddress]{revtex4-1}
\usepackage{}

\usepackage{mathrsfs}
\usepackage{amssymb}  

\usepackage{graphicx}
\usepackage{dcolumn}
\usepackage{bm}
\usepackage{ifthen}
\usepackage{booktabs}
\usepackage{amsmath}

\begin{document}


\title{Quantitative verification of the Kibble-Zurek mechanism in quantum non-equilibrium dynamics}



\author{Xiao-Ye Xu}
\affiliation{Key Laboratory of Quantum Information, University of Science and Technology of China, CAS, Hefei, 230026, People's Republic of China}

\author{Yong-Jian Han}
\email{smhan@ustc.edu.cn}
\affiliation{Key Laboratory of Quantum Information, University of Science and Technology of China, CAS, Hefei, 230026, People's Republic of China}

\author{Kai Sun}
\affiliation{Key Laboratory of Quantum Information, University of Science and Technology of China, CAS, Hefei, 230026, People's Republic of China}

\author{Jin-Shi Xu}
\affiliation{Key Laboratory of Quantum Information, University of Science and Technology of China, CAS, Hefei, 230026, People's Republic of China}

\author{Jian-Shun Tang}
\affiliation{Key Laboratory of Quantum Information, University of Science and Technology of China, CAS, Hefei, 230026, People's Republic of China}

\author{Chuan-Feng Li}
\email{cfli@ustc.edu.cn}
\affiliation{Key Laboratory of Quantum Information, University of Science and Technology of China, CAS, Hefei, 230026, People's Republic of China}

\author{Guang-Can Guo}
\affiliation{Key Laboratory of Quantum Information, University of Science and Technology of China, CAS, Hefei, 230026, People's Republic of China}

\date{\today}

\begin{abstract}
The Kibble-Zurek mechanism (KZM) captures the key physics in the non-equilibrium dynamics of second-order phase transitions, and accurately predict the density of the topological defects formed in this process. However, despite much effort, the veracity of the central prediction of KZM, i.e., the scaling of the density production and the transit rate, is still an open question. Here, we performed an experiment, based on a nine-stage optical interferometer with an overall fidelity up to 0.975$\pm$0.008, that directly supports the central prediction of KZM in quantum non-equilibrium dynamics. In addition, our work has significantly upgraded the number of stages of the optical interferometer to nine with a high fidelity, this technique can also help to push forward the linear optical quantum simulation and computation.
\end{abstract}


\maketitle

In the early universe after the "Big Bang", cosmological phase transitions occurred with the expansion and cooling of the universe, and the symmetry of the vacuum was broken. The new vacuums were chosen locally, within space-like regions, resulting in topological defects\cite{Kibble1976,Kibble1980}. The initial density of the topological defect is extremely interesting, and a rough limit of this density can be estimated by the light-cone causality; however, the exact density is not easy to determine. Zurek suggested that this cosmological mechanism can be observed in condensed matter systems in a laboratory \cite{Zurek1985,Zurek1993,Zurek1996}. For example, a pressure quench drives liquid $^4$He from a normal phase to a superfluid phase at a finite rate, which leaves behind vortex lines.

In a condensed matter system, the speed limit of the light is less useful to estimate the density of defects. However, the density of topological defects can be predicted for second order phase transitions due to the divergence of the relaxation time $\tau$ (which characterizes the time required for the order parameter to relax to its equilibrium value when the parameter has been perturbed) and the healing length (which characterizes the length over which the order parameter will return to the equilibrium value when disturbed) near the critical point. As a result of the divergence, every such transition, traversed at a finite rate, is inevitably a non-equilibrium dynamical process. The whole system can not catch up, and the symmetry will be broken with some topological defects \cite{Zurek1985,Zurek1993,Zurek1996}. Therefore, the Kibble-Zurek mechanism (KZM) provides a theoretical framework with which to describe the non-equilibrium dynamics of the symmetry broken in the second order transition \cite{Kibble1976,Kibble1980,Zurek1985,Zurek1993,Zurek1996}.

\begin{figure*}
\begin{center}
\includegraphics[width=6.5in]{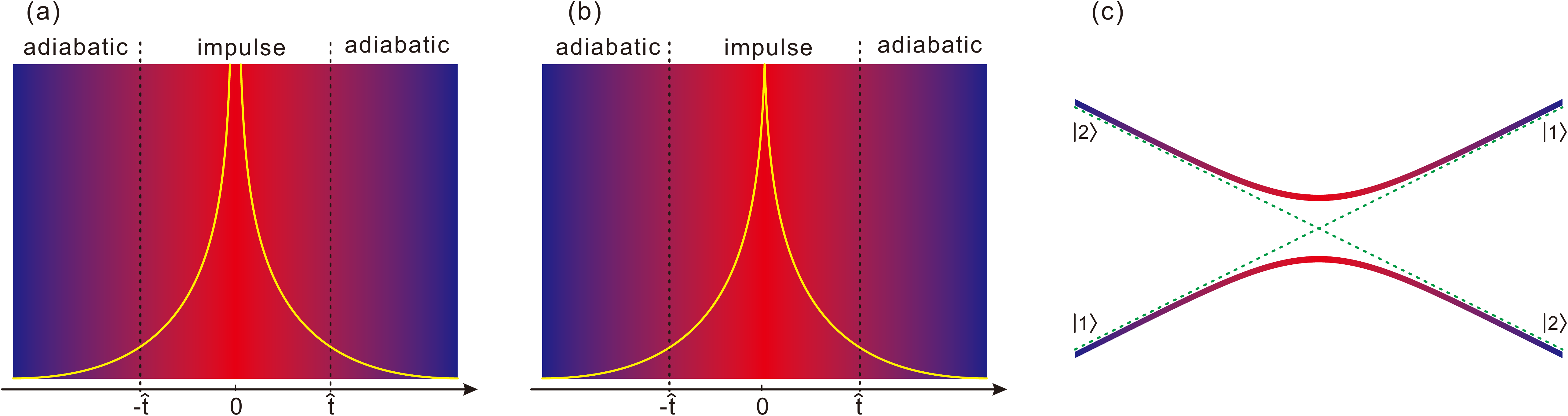}
\end{center}
\caption{(color online) The comparison of the features in the KZM and the quantum non-equilibrium dynamics of the LZ model. (a) shows the relaxation time scale $\tau$ in the KZM, and (b) shows the inverse of the energy gap in the LZ model. (c) gives the structure of the energy levels in the LZ model.} \label{fig:compare}
\end{figure*}

The central prediction of the KZM is that the density of the topological defects in the second order phase transition should scale with the transit rate \cite{Zurek1985,Zurek1993,Zurek1996}. To be more specific, consider the phase transition in liquid $^4$He driven by the pressure (denoted here by $t$) and suppose that the critical point is located at $t = 0$. The entire dynamic process can be divided into three parts, the adiabatic, impulse and adiabatic regions(which are shown in Fig.\ref{fig:compare}(a)), according the relaxation time scale $\tau$. Between $-\hat{t}$ and $\hat{t}$, the system can not adjust in time to follow the new thermodynamic conditions and is effectively frozen. The freeze-out time $\hat{t}$ is determined by the Zurek equation \cite{Zurek1985,Zurek1993,Zurek1996}: $\tau(\hat{t}) = \hat{t}$. The relaxation time $\tau$ with a general parameter $t$ can be determined experimentally as  $\tau=\tau_0/|\varepsilon|$ for the case of liquid $^4$He (where $\tau_0$ is a constant, and $\varepsilon$ with $\varepsilon(t=0)=0$ is called the relative temperature, which is used to characterize the distance between the system and the critical point). In addition, $\varepsilon$ can be related to the transit rate $\tau_Q$ in the liquid $^4$He case as $\varepsilon(t)=t/\tau_Q$. Therefore, the freeze-out time can be determinged with the Zurek equation as $\hat{t}=\sqrt{\tau_0\tau_Q}$. As analyzed in \cite{Zurek1985,Zurek1993,Zurek1996}, the density of the topological defects produced in the non-equilibrium dynamics of the transition can be predicted with the knowledge of $\hat{t}$ without solving the dynamical equations.

For the ubiquity of the second order transition and non-equilibrium dynamics, it is very interesting to verify the KZM in the laboratory. Though many efforts have been made\cite{Hendry1994,Ruutu1996,Bauerle1996,Chuang1991,Bowick1994, Ducci1999,Carmi2000,Maniv2003,Sadler2006,Weiler2008}, the central prediction of the KZM has still not been observed clearly\cite{Arnab2012}. Because the KZM is a framework for non-equilibrium dynamics and the anticrossing between the ground state and the first excitation state of a generic second order phase transition\cite{sachdev}, Damski \cite{Damski2005} presents a simple scheme to support the KZM in a quantum non-equilibrium dynamical system - the Landau-Zener (LZ) model with the time dependent Hamiltonian $H=(\omega_0\sigma_x+\Delta t\sigma_z)/2$ \cite{Landau,Zener,Damski2005}. By introducing correspondence between the LZ model and the second order phase transition, such as, $\tau_Q:=\omega_0/\Delta$, $\tau_0:=1/\omega_0$ and the density of defects $D_n:=|\langle\Psi|1\rangle|^2$, and based on the framework of the KZM, Damski provides a direct prediction of the transition probability long after the critical point, which corresponds to the density of the topological defects in the second order phase transition (The details are shown in Fig.\ref{fig:compare}, and the procedure for comparison is given in the supplementary information). Suppose the dynamics start at $t = 0$, the transition probability at $t \gg 0$ under the KZM should then be
\begin{equation}
\mathcal {D}_n = 0.5(1-\sqrt{1-2/\mathcal{P}(x_\alpha)}),
\label{1}
\end{equation}
where $x_\alpha = \alpha\frac{\tau_Q}{\tau_0}$ and $\mathcal{P}(x_\alpha) = x_\alpha^2 + x_\alpha\sqrt{x_\alpha^2+4}+2$ (see the supplementary information and \cite{Damski2005}). Eq.\ref{1} is a direct result of Zurek's equation and reflects the key quantitative aspects of KZM, i.e., the scaling rule, in this quantum dynamical process.

To verify Eq.\ref{1}, which is predicted by the KZM, we need to realize a quantum dynamical process, which is driven by a time-dependent Hamiltonian. The dynamical process is not easily implemented in linear optics. The traditional methods reported as quantum simulations in linear optics\cite{XiaoSong2011}, which require the prior information of the state and can be viewed as a state preparation process, can not be used in this situation. Although the information of the state of the simple LZ model can be calculated by solving the dynamical equations, the traditional method is not a dynamical process and can not be used to verify the KZM. In the dynamical process, we only have the information of the initial state and the time-dependent Hamiltonian (or the corresponding unitary operator), and have no prior information about the middle state and the final state. We first experimentally implement this dynamical process in linear optics. We divide the dynamical process into many segments due to the difficulty of tuning the Hamiltonian continuously; in each segment (a step in the experiment), the Hamiltonian has a fixed parameter according to a different time. This process is a truly quantum simulation in linear optics and our method has been used to realize an imaginary time evolution and demonstrate the algorithmic quantum cooling \cite{JinShi}. Here, we extend the process to a real time evolution with more stages. With this method, we observe the evolution of the LZ model, and by comparing the experimentally measured transition probability and the prediction of Eq.\ref{1}, the first experimental observation of the KZM in a non-equilibrium quantum dynamical system is presented.

\begin{figure*}
\begin{center}
\includegraphics[width=6.5in]{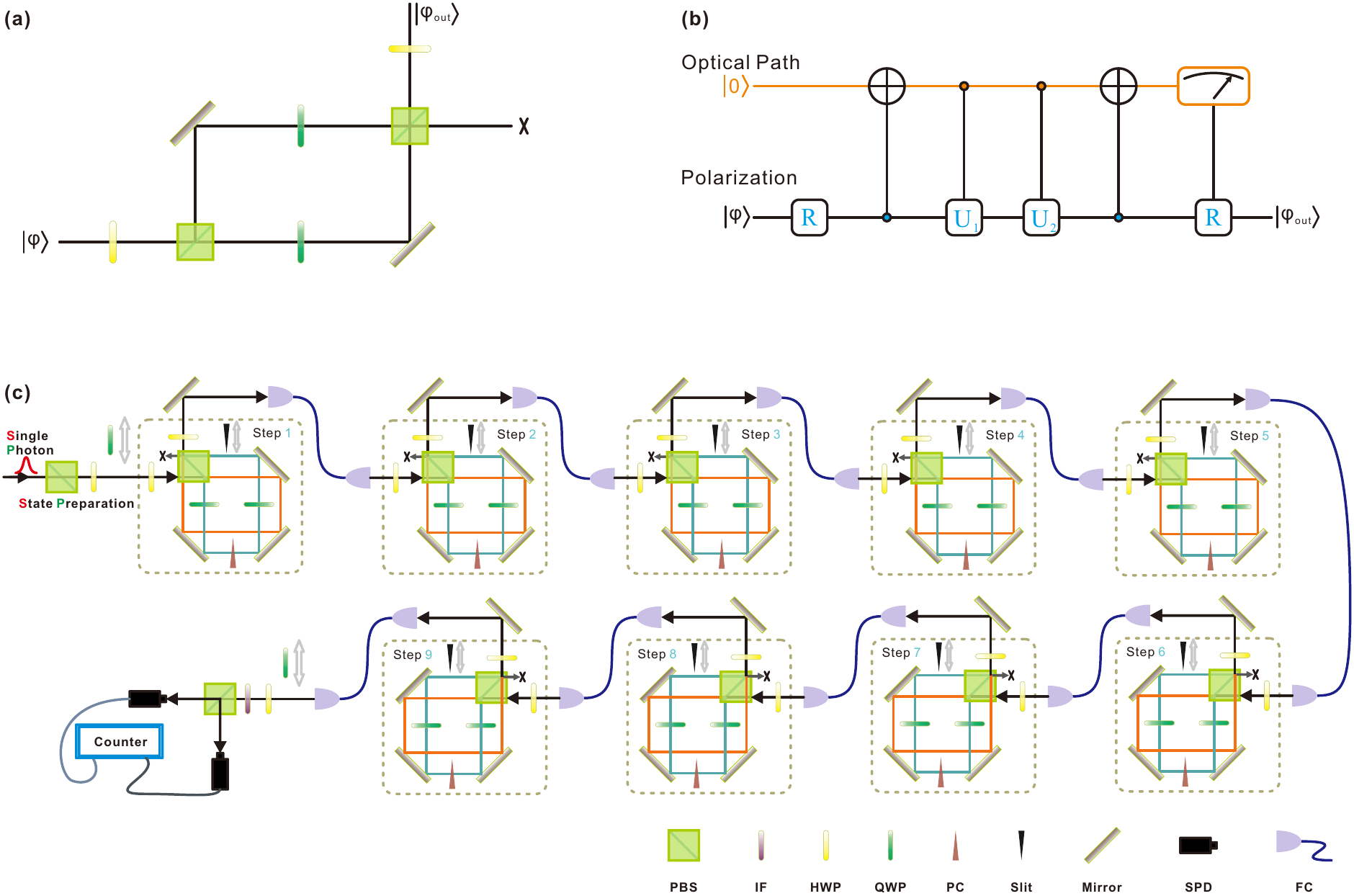}
\end{center}
\caption{(color online) (a) A standard MZI, whose logic diagram is shown in (b). (c) provides the experimental details. FC stands for fiber collector.} \label{fig:setup}
\end{figure*}

The basic module in our experiment, shown in Fig.\ref{fig:setup}(a), is a polarization based interferometer, known as the Mach-Zender interferometer (MZI), with some local operations that are realized by wave plates inside and outside of the device. The MZI is one of the most basic elements in a linear optical quantum computation. It has been noted that, in principle, the quantum logic in a quantum computation can be simulated universally by an MZI \cite{PhysRevA.57.R1477}. Figure.\ref{fig:setup}(b) shows the logic diagram of this structure, and the overall operation is then given by $\frac{1}{2\sqrt{2}}(I-i\sigma_x\sin{4\alpha}\cos{2\gamma}-i\sigma_z\cos{4\alpha}\cos{(2\gamma)})$ where $\alpha(\gamma)$ represents the angle of the optical axis of the HWPs (QWPs) and $\sigma_x$ and $\sigma_z$ are two Pauli operators. In digital quantum simulators, to obtain the final state of a quantum system governed by a time-dependent Hamiltonian $H_{sys}(t)$ for a time interval $T$, we divide the entire process into $n$ steps. In each step, with a time interval $\tau=T/n$, $H_{sys}(t)$ will not change significantly and can be replaced with a time-independent Hamiltonian $H(t_k)$, where $k$ is the index of the step. The time $t_k$ can be at any time position of the step $k$. For convenience, it is chosen here at the middle of the step. For a initial state $|\phi\rangle$, the final state can then be approximated by $\prod_k e^{-i\hbar H(t_k)\tau}|\phi\rangle$. The only remaining task is to realize each step's evolution and, step by step, to then obtain the approximate final state. For each step's evolution, we can replace $e^{-i\hbar H(t_k)\tau}$, approximately, with $I-i\hbar H(t_k)\tau$, where $I$ is the identity operator. For the Hamiltonian in the LZ model, the approximate local time evolution operator is $U_k = I-i\sigma_x\theta-i\sigma_z\theta\epsilon_k$,
where $\theta=\frac{\omega_0\tau}{2\hbar}$ and $ \epsilon_k = \frac{\Delta t_k}{\omega_0}$. Comparing this operator with the operator formula given above, we can obtain the relevant value of $\alpha(\gamma)$ and adjust the Hamiltonian in each step by changing the angles of the wave plates.



Due to the intrinsic phase instability in the standard MZI, it is impossible to build an interferometer with complicated structures and multiple stages. That is to say, we can not extract any practical quantum logic based on the standard MZI. Scientists have developed an increasing number of schemes to overcome this disadvantage \cite{Broome,Shadbolt}. Thus far, these methods have provided no more than five stages. Here, shown in Figure.\ref{fig:setup}(c), we upgrade the number of stages in optical interferometer to nine based on another alternative device, the Sagnac-type interferometer, which has been used to realize a partial measurement \cite{XiaoYe2011} and has been developed to three stages to demonstrate the algorithmic quantum cooling \cite{JinShi}. In our experiment, the visibility of every stage is over 1000:1 and the overall fidelity is measured to be 0.975$\pm$0.008, which is the highest known value that provides sufficient intrinsic stability to perform small-scale quantum computations. The characterization of nine-stage interferometer is given in detail in the supplementary section.

\begin{figure}
\begin{center}
\includegraphics[width=3.3in]{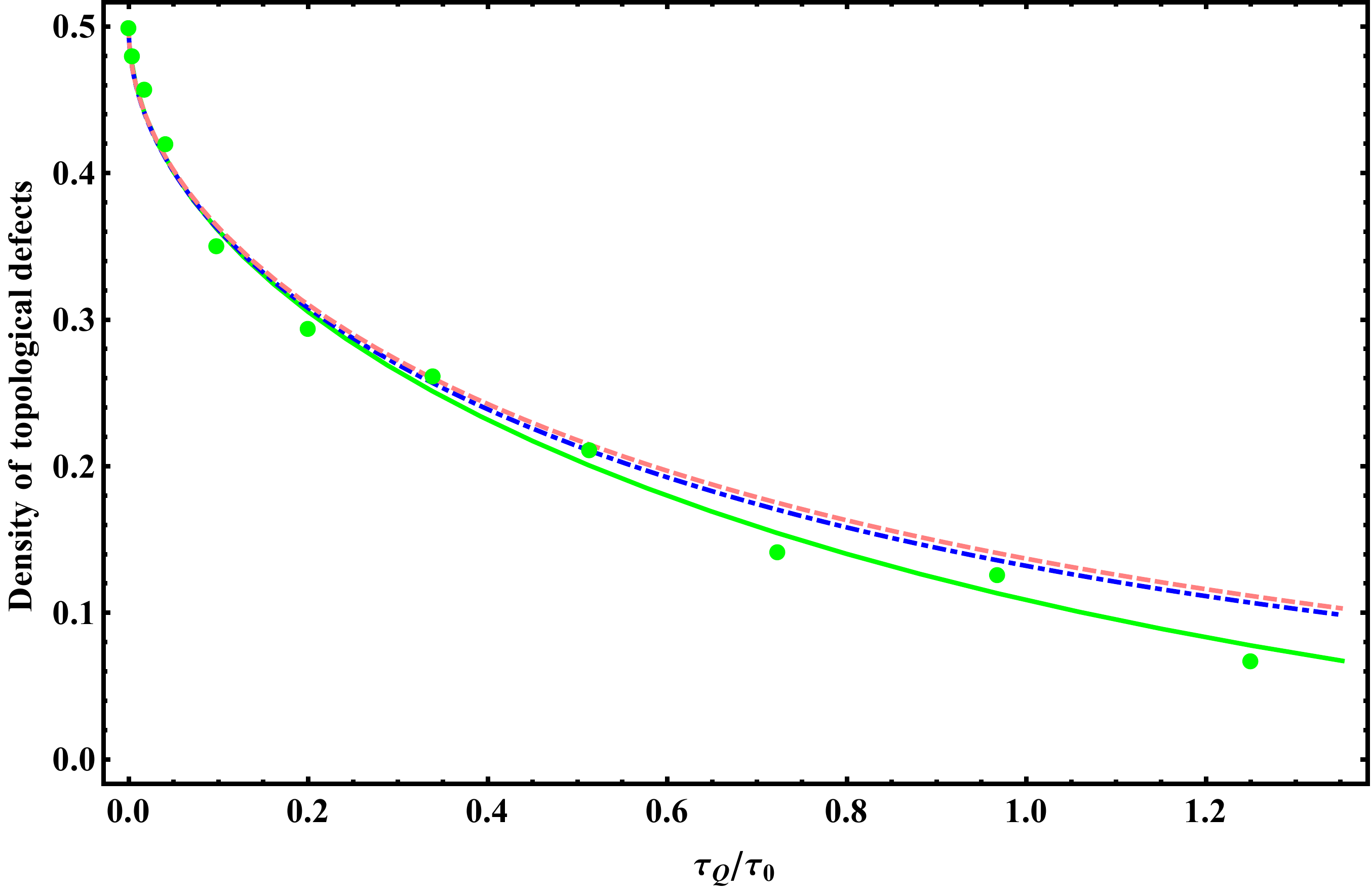}
\end{center}
\caption{(Color online) The experimental results compared with the theoretical prediction. The blue dash-dotted  line is the theory prediction from the Kibble-Zurek mechanism with $\alpha = 0.8$. The experimental results are shown by the points, and the green line shows the numerical solution of our experimental system. The pink dashed line gives the analytical solution with the same parameters.}
\label{fig:kzm}
\end{figure}

The main purpose of this work is to show the Kibble-Zurek mechanism in the non-equilibrium dynamics of the LZ model. First, we choose the parameters $\Delta = 5$ and $\tau = 2$ and perform a simulation of the evolution of the LZ model (The results are shown in the supplementary information). Due to the limitation of the number of stages, we have clearly observed one and a half periods in the evolution of the LZ model. Then, to achieve the main goal, we change the parameter $\omega_0$ from 0.05 to 2.5, which enables us to observe the LZ dynamics from the near equilibrium case to the non-equilibrium case. We compared the density of the $H$ polarization after the evolution by the nine-stage interferometer and the theoretical prediction by Eq.\ref{1}; these values are shown in Figure.\ref{fig:kzm}. When the value of $\tau_Q/\tau_0$ is small, which is the case in the non-equilibrium region, we find that the density of the upper lever measured in the experiment is closely equivalent to the density of the topological defects predicted by the KZM (shown by blue dash-dotted line). The green line gives the numerical solution of our experimental system, which deviates from the KZM prediction in the large region of $\tau_Q/\tau_0$. There are two reasons for this deviation. First, the limited number of stages in our experiment (which we believe to be the maximum number of stages achieved thus far) means the evolution has not completely finished (shown in the supplementary section); second, the system has deviated from the non-equilibrium region, which can be confirmed by the exact analytical solution of the LZ model, shown by the pink dashed line in Figure.\ref{fig:kzm}.

In conclusion, a nine-stage interferometer has been built with a high visibility and intrinsic stability. By using this device, we observe the dynamical evolution of LZ model. As the main objective of this work, we change the quench time scale in the LZ model and measure the final density of the upper level which corresponds to the density of the topological defects in the KZM. We then provide an experimental confirmation that the KZM emerges from the simplest quantum non-equilibrium dynamics. We hope that these results will be helpful in understanding the KZM in quantum non-equilibrium dynamical system and in revising the simplest system from another perspective. Our platform will be helpful in quantum walks and quantum computations. When we completed our experiment and prepared this manuscript, we noticed that K. Pyka et al, \cite{Pyka}, reported an experimental result to demonstrate the scaling of the number of topological defects with the transition rate in ion trap.

\section*{Methods}
The experimental setup is shown in Figure.\ref{fig:setup}(c). Coherent pulses from a Ti:sapphire laser with the center wavelength mode locked at 800\,nm (with a 130\,fs pulse width and a 76\,MHz repetition rate) are adequately attenuated to the single-photon level, which is less than 0.007 photons in each pulse, and coupled to a nine-stage polarization based interferometer via a single mode fiber (SMF).  This resource has been employed to demonstrate quantum walks \cite{Schreiber} and has also been widely used in practical quantum key distributions \cite{Zhao2006}. The initial polarization state is prepared by using a polarized beam splitter (PBS) followed by a half wave plate (HWP). Each step of the evolution process is realized by a module composed of a polarization-dependent Sagnac interferometer. By using two quarter-wave plates (QWPs) independently in each arm and two HWPs (one in front of and the other after the interferometer), we can adjust the evolution parameters; this process is described in detail in the main text. A phase compensator (PC) is inserted in one arm to correct the phase error and there is an adjustable slit in the other arm for calibration.  A 1\,m long SMF is used as the coupler between two independent interferometers and also acts as a spatial filter. When the evolution is finished, the photons are collected by an SMF and guided to the polarization analysis module, which is consists of an HWP, a Wollaston prism and two single photon detectors (SPDs). One interference filter (IF) with a 1\,nm band width and an 800\,nm center is introduced before the Wollaston prism to increase the coherence length. Two QWPs, one after the state preparation and the other before the polarization analysis module, are inserted to perform the process tomography.

\textbf{About the errors:}
In our work, the errors are primarily due to three factors: The first error is the errors of the wave plates. There are more than 30 wave plates in the setup, and each one has an angle error of approximately 0.1 degree: the total error can be simulated numerically and is found to be 0.011 by the Monte Carlo method. The second error is the shot noise. In our experiment, the total number of photons is over $5\times10^5$: according to the numerical calculation, the error due to the shot noise is much smaller than the angle error of the wave plates. The last error is due to the power fluctuation of the photon source: here, we employ two detectors to detect all of the outputs and overcome this disadvantage by renormalization.

\section*{Acknowledgement}
This work is supported by the CAS. We are grateful to the following funding sources: the National Basic Research Program of China (Grant No. 2011CB9212000), National Natural Science Foundation of China (Grant Nos. 60921091, 11274289,11274297,11004185), the Fundamental Research Funds for the Central Universities (Grant Nos. WK2470000004, WK2470000006, WK2470000007, WK2030020019) and NSFC11105135.
\section*{Supplement for quantitative verification of the Kibble-Zurek mechanism in quantum non-equilibrium dynamics}
\subsection{Landau-Zener model as the simplest version supporting the Kibble-Zurek Mechanism}
The LZ model is a well understood quantum dynamical model for the relatively simple form of the time-dependent Hamiltonian  $H_{LZ}=0.5(\Delta t\sigma_z+\omega_0\sigma_x)$, which has an analytical solution in theory. In this model, the so called LZ transition acts as a result of the tuning, and the transition probability depends on both the energy gap in the critical point, $\omega_0$, and the velocity of the tuning of the LZ Hamiltonian, $\Delta$. The former is decided by the quantum system itself; increasing the gap decreases the probability. The latter is related to the speed of the tuning of the external field. These two parameters determine the instant energy gap, $\sqrt{\omega_0^2+(\Delta t)^2}$. According to the adiabatic theorem, when the inverse of the gap is small enough, the system can catch up with the changes of the tuning and remains in the instant ground state if the system began from a ground state at $t\rightarrow -\infty$; that is to say, the system undergoes an adiabatic evolution. Therefore, as Damski claimed, it is naturally suggested that the inverse of the instant energy gap acts as a quantum mechanical equivalent of the relaxation time scale $\tau$ introduced above, which indicates how much time the system needs to adjust to new thermodynamic conditions \cite{Damski2005}. As the direct results of the dynamics, the transition probabilities are suggested to identify the density of topological defects - the main concern in the KZM. Then, as a direct suggestion, the other two fundamental parameters in the KZM, the relative temperature $\epsilon$ and the quench time scale as $\tau_Q$, can be defined with $\Delta t/\omega_0$ and $\omega_0/\Delta$, respectively, where $\omega_0 = 1/\tau_0$.

In the KZM, all the predictions of the density of the topological defects can be made with the knowledge of the frozen-out time $\hat{t}$, which satisfies Zurek's equation. According to the analogy presented above, this result means that the transition probability in the LZ model can be obtained from the knowledge of $\hat{t}$. With simple mathematical calculations, Damski presented two cases to show that the transition probability, which corresponds to the density of the topological defects, can be predicted directly and perfectly from the modified version of Zurek's equation by introducing a free parameter $\alpha$ to the equation. Considering the case when the time evolution starts from a ground state at the anti-crossing center, the density of the topological defects is suggested to be \cite{Damski2005},
\begin{equation}
\mathcal {D}_n = 0.5(1-\sqrt{1-2/\mathcal{P}(x_\alpha)}),
\end{equation}
where $x_\alpha = \alpha\frac{\tau_Q}{\tau_0}$ and $\mathcal{P}(x_\alpha) = x_\alpha^2 + x_\alpha\sqrt{x_\alpha^2+4}+2$.

Due to the simple form of the Hamiltonian, the LZ model is experimentally obtainable and the transition has been observed in many quantum systems, from Rydeberg atoms \cite{Baruch,Yoakum} to quantum dots contacts \cite{Gorelik}, and recently extended to mesoscopic superconducting Josephson devices \cite{Sillanp}, ultracold molecules \cite{Mark}, optical lattices \cite{Kling} and the NV center \cite{Huang}. There are no technological challenges in controlling the level structure in LZ dynamics in most of these systems and more complicated dynamics based on the LZ model have been presented (see \cite{Schevchenko} for a review). Therefore, on the technological side, the verification of Eq.\ref{1} and the observation of the KZM in such a two-level dynamical system are possible.

\subsection{Characterization of the interferometer}
\begin{figure*}
\begin{center}
\includegraphics[width=5.5in]{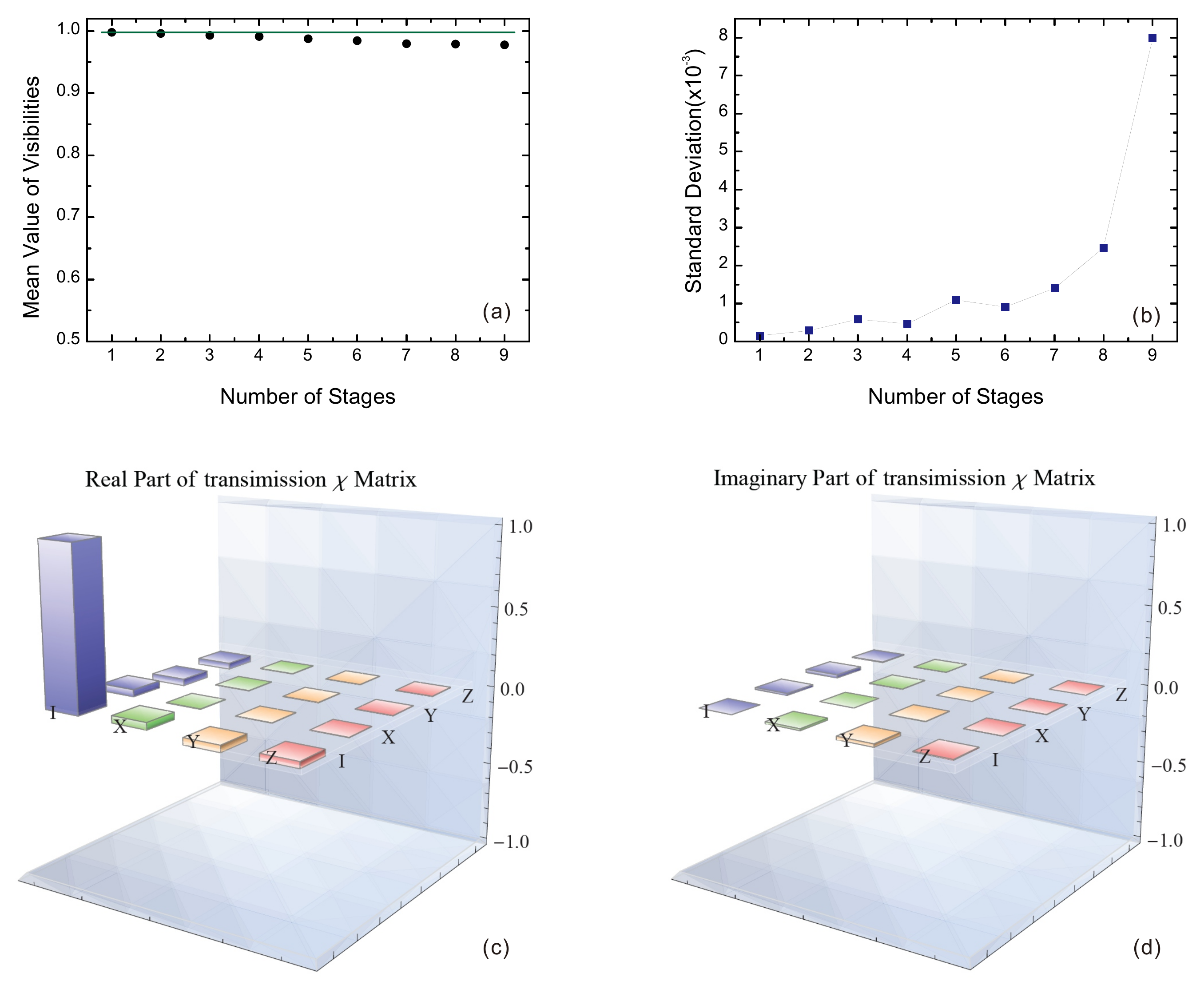}
\end{center}
\caption{(Color online) The experimental results from tests of the visibility and stability of the nine-stage interferometer. Figure (a) gives the mean values of visibility from one- to nine-stages, where the error bars are negligible. The data were collected for more than one hour. Figure (b) shows the standard deviation of the data, which is an indicator of the stability. (c) and (d) show the real and imaginary parts of the $\chi$ matrix of the nine-stage interferometer achieved from the quantum process tomography \cite{JModOpt.44.2455}.}
\label{visibility}
\end{figure*}
Two of the most important parameters of an interferometer are the visibility and stability.  Ultra-high visibility interferometers have been used in various fundamental physics, such as the laser interferometer gravitational-wave observatory \cite{Science.256.325}. Usually, the visibility is a direct indicator of the operation fidelity. The stability is an indicator of decoherence; a higher stability means less decoherence before the operations are finished. Here, we test our setup by measuring both the visibility and the stability, which are shown in Figure.\ref{visibility}. The integration time of a single measurement is 1\,s, and we monitor the phase stability for more than one hour. With the help of spectrum and spatial filter, the visibility is higher than $0.998$ for each stage. Then we connect the stages one by one with a 1\'m long single-mode fiber. The measured visibility decreases slowly to $0.975\pm0.008$ as the stages increase to nine, which fits the theory prediction of $0.998^9$ very well considering the error. We calculate the standard deviation to investigate the stability in one hour. From the results shown in Figure.\ref{visibility}(b), we can see that the stability worsens as the number of stages increases. However, due to the intrinsically stable feature of the Sagnac type architecture, even a nine-stage device is sufficiently stable to conduct the experiment. This finding is also confirmed with the $\chi$ matrix ((c) and (d) in Fig.\ref{visibility}), which is reconstructed by the quantum process tomography. For an ideal case, the matrix equals $I$. Here, the measured fidelity is $0.994\pm0.011$.

\subsection{Observation of the evolution in LZ model}
\begin{figure}
\begin{center}
\includegraphics[width=3.3in]{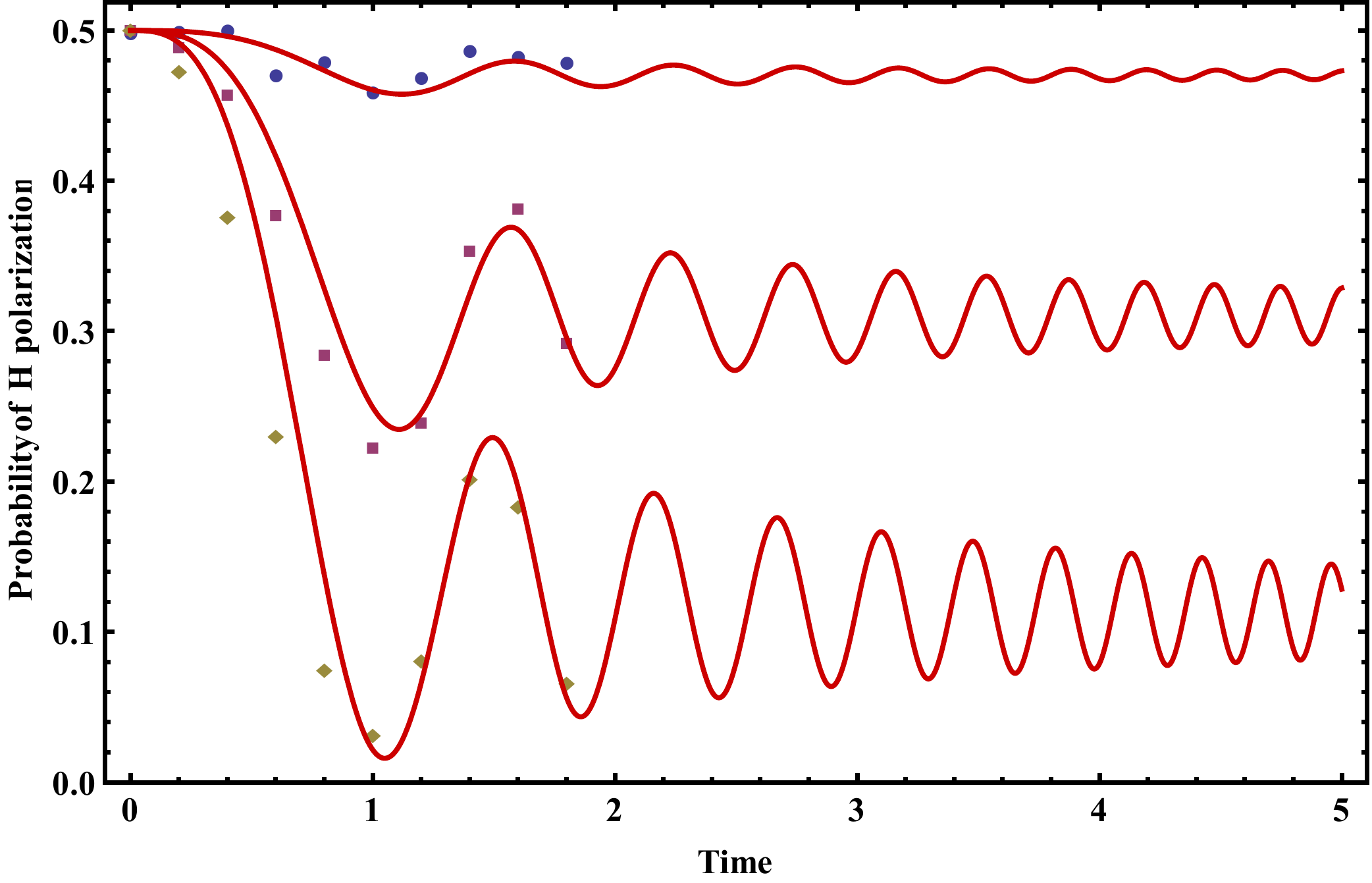}
\end{center}
\caption{(Color online) An experimental demonstration of the evolution of a two-energy-level quantum system governed by the Landau-Zener Hamiltonian. From the upper to lower panels, we compare the different features for different values of $\omega_0^2/\Delta$.}
\label{fig:lz}
\end{figure}

With the help of this nine-stage interferometer, we perform the simulation of the evolution of the LZ model, and the results are shown in Figure.\ref{fig:lz}. Here, we choose the parameters of $\Delta = 5$ and $\tau = 2$ for each step. Due to the limitation on the number of stages, one and a half periods in the evolution of the LZ model can be observed clearly. The points agree with the theoretical predictions very well.

\bibliography{reference}

\end{document}